\documentclass[11pt]{article}

\usepackage{amsfonts}
\usepackage{epsfig}
\usepackage{latexsym}
\usepackage{amsmath}
\usepackage{amssymb}
\usepackage{float}

\def\bequ{\begin{equation}}
\def\eequ{\end{equation}}
\def\barr{\begin{array}}
\def\earr{\end{array}}

\def\ben{\begin{equation}}
\def\een{\end{equation}}
\def\bena{\begin{eqnarray}}
\def\eena{\end{eqnarray}}

\def\b1{e^0}

\newcommand{\be}{\begin{equation}}
\newcommand{\ee}{\end{equation}}
\def\bea{\begin{eqnarray}}
\def\eea{\end{eqnarray}}

\def\be{\begin{equation}}
\def\ee{\end{equation}}
\def\bea{\begin{eqnarray}}
\def\eea{\end{eqnarray}}

\def\lesssim{\mathrel{\hbox{\rlap{\hbox{\lower4pt\hbox{$\sim$}}}\hbox{$<$}}}}
\def\gtrsim{\mathrel{\hbox{\rlap{\hbox{\lower4pt\hbox{$\sim$}}}\hbox{$>$}}}}

\begin{document}
\title{{\LARGE \bf{Relativistic Euler's 3-body problem, \\ optical geometry and the golden ratio}}}
\author{
Fl\'avio S. Coelho\footnote{flavio@physics.org}~  and 
Carlos A. R. Herdeiro\footnote{crherdei@fc.up.pt}
\\
\\ {\em Departamento de F\'\i sica e Centro de F\'\i sica do Porto}
\\ {\em Faculdade de Ci\^encias da Universidade do Porto}
\\ {\em Rua do Campo Alegre, 687,  4169-007 Porto, Portugal}}

\date{September 2009}       
 \maketitle

\begin{abstract}
A Weyl solution describing two Schwarzschild black holes is considered. We focus on the $\mathbb{Z}_2$ invariant solution, with ADM mass $M_{ADM}=2M_K$, where $M_K$ is the Komar mass of each black hole. For this solution the set of fixed points of the discrete symmetry is a totally geodesic sub-manifold. The existence and radii of circular photon orbits in this sub-manifold are studied, as functions of the distance $2L$ between the two black holes. For $L\rightarrow 0$ there are two such orbits, corresponding to $r=3M_{ADM}$ and $r=2M_{ADM}$ in Schwarzschild coordinates.  As the distance increases, it is shown that the two photon orbits approach one another and merge when $M_K=\varphi L$, where $\varphi$ is the golden ratio. Beyond this distance there exist no circular photon orbits.
The two null orbits delimit a forbidden band for time-like circular orbits, which is interpreted in terms of optical geometry. For large $L$, time-like circular orbits are allowed everywhere, as in the analogous Newtonian problem. The analysis is generalised by considering a $\mathbb{Z}_2$ invariant Weyl solution with an array of $N$ black holes and also by charging the black holes, which connects the Weyl solution to a Majumdar-Papapetrou spacetime.
\end{abstract}



\section{Introduction}
Euler's 3-body problem is a soluble special case of the general 3-body problem, in which a test particle moves in the gravitational field of two gravitational sources - point masses - which are fixed in space. It is Liouville integrable since, besides the energy and one component of the angular momentum, there is a third constant of the motion (see e.g. \cite{Coulson,Bell}; see \cite{Gibbons:2006mi} for a hyperbolic space version of the problem which is still integrable). A natural relativistic generalisation of this problem is to consider the motion of test particles in a Weyl solution describing two Schwarzschild black holes, which are kept in equilibrium due to a strut in between them. For this problem there is no known analogue to Euler's constant (see \cite{Will:2008ys} for a discussion about this point); equivalently, there is no known separability for the geodesic equations. Generic geodesics have, therefore, to be studied numerically. For the special case in which the two black holes have equal mass, there is, however, a 1+2 dimensional totally geodesic sub-manifold in which geodesics are Liouville integrable. In particular it is straightforward to derive all circular causal orbits.

Regular (on and outside an event horizon) static black holes in vacuum general relativity do not admit causal circular geodesics below a certain critical radius, which lies outside the event horizon. Below this radius, the angular momentum term in the radial equation of motion, usually centrifugal (with respect to the black hole), becomes centripetal, making equilibrium impossible. This change in character of the angular momentum term has a geometric interpretation in terms of optical geometry \cite{Abram}: define ``outwards" as the direction in which fixed points sets of the optical geometry's isometry group (spheres) increase their proper size; then, the angular momentum term is always ``outwards". Similar considerations may be made about stationary black holes. However, in this case the optical geometry is a Randers-Finsler rather than Riemannian geometry, making the interpretation more subtle \cite{Gibbons:2008zi}.

In this paper we show that the optical geometry of the aforementioned relativistic Euler's 3-body problem is richer than the standard one of the Schwarzschild black hole. To be concrete let the two black holes in 
the Weyl solution considered have equal mass $M_K$ and separation $2L$ between them. This choice of masses implies that there is a discrete $\mathbb{Z}_2$ symmetry of the solution. The set of fixed points of this discrete symmetry is a totally geodesic sub-manifold, $\mathcal{N}$. The optical geometry of this sub-manifold then has the following feature: for sufficiently small but non-zero $L$ the notion of  ``outwards", defined as above, changes continuously twice, rather than just once. This means that, on $\mathcal{N}$, there is an annulus-like region in which "outwards" means a decreasing radial Weyl canonical coordinate. The boundaries of this annular region are the loci of two circular photon orbits, and the region itself is a forbidden band for time-like circular geodesics, which are allowed everywhere outside this band. As $L$ increases, the two boundaries of the annular region approach one another and coalesce when $M_K=\varphi L$, where $\varphi$ is the golden ratio! For larger $L$, $\mathcal{N}$ admits no circular photon orbits and time-like circular geodesics are allowed everywhere, just as in the $\mathbb{Z}_2$ symmetric Euler's 3-body problem.

Considering a Weyl solution with an array of $N$, rather than two, black holes along a line, still possessing a $\mathbb{Z}_2$ discrete symmetry, leads to two cases. For $N$ even (odd), we essentially recover the optical geometry of the $N=2$ ($N=1$) case. We shall also consider charged black holes in Einstein-Maxwell theory, by using a solution generating technique \cite{Azuma} that allows the introduction of a charge parameter in Weyl solutions. For black holes, the charge parameter is the ratio of charge to mass, which is the same for all the black holes in the solution. In this case, as we increase the charge for fixed $L, M_K$, the proper area of the forbidden band increases; for $L=0$, this area diverges as we reach extremality, as a consequence of the infinite `throat' developed by extremal black holes. Keeping the charge and $M_K$ fixed, the area of the forbidden band decreases as $L$ is increased, just as for the uncharged case, becoming zero at some maximum value of $L$. 

This paper is organised as follows. After a brief discussion of the Newtonian Euler's 3-body problem in section 2, we describe in section 3.1 the Weyl solutions that shall be used in the relativistic version of the problem, as well as the circular null and time-like orbits. In section 3.2 the charged case is considered. In section 4 the interpretation of the forbidden band for time-like circular orbits is given in terms of optical geometry. In section 5 we briefly discuss the case of multiple black holes. We close with some final remarks.

\section{The Newtonian version}
\begin{figure}[h!]
\centering\includegraphics[height=3.1in]{{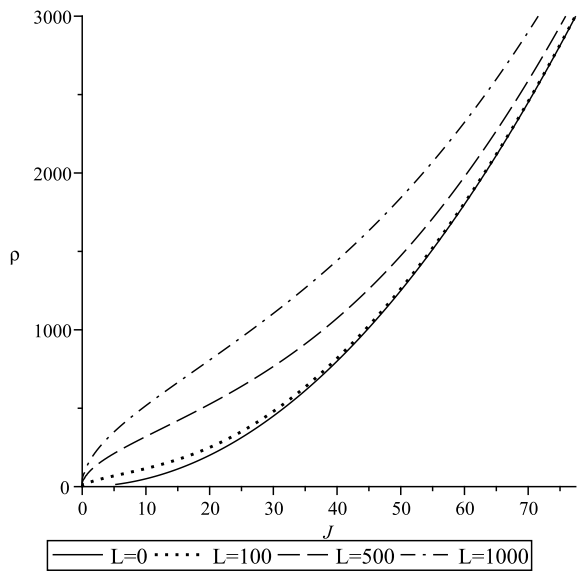}}
\caption{Radial distance, $\rho$, of circular orbits in the symmetry plane of the two-centre Kepler problem, as a function of their angular momentum, $J$. The two equal masses, which we set to unity, have separation $2L$, and various values of $L$ have been used. Observe that circular orbits exist for any $\rho$.}
\label{newtonian}
\end{figure}

In Euler's 3-body problem, a.k.a. the two-centre Kepler problem (see e.g. \cite{jose}), with both masses equal to $M$, the motion of a test particle moving in the symmetry plane between the two masses obeys
\bequ
\left(\frac{d\rho}{dt}\right)^2=2E-V(\rho) \ , \qquad V(\rho)=\frac{J^2}{\rho^2}-\frac{4M}{\sqrt{\rho^2+L^2}} \ , 
\label{radequ}
\eequ
where $\rho$ is a radial cylindrical coordinate, $E,J$ are the energy and angular momentum per unit mass and $L$ is the distance from either mass to the symmetry plane. We shall use geometrised units throughout. Circular orbits are seen by extremising the potential. They obey
\bequ
\frac{J^2}{2M}=\frac{\rho^4}{(\rho^2+L^2)^{3/2}} \ . \eequ
This equation has solution for any $\rho$, given $L$ and $M$. The angular momentum $J$ goes to zero (infinity) as $\rho\rightarrow 0$ ($\rho\rightarrow \infty$) -  Fig. \ref{newtonian}. Thus circular orbits exist for any radial distance in the symmetry plane. This is therefore the behaviour we expect in the relativistic version of the problem for large $L/M$, and indeed it is the behaviour we shall find. The behaviour for small $L/M$ will, however, be quite different.

In the relativistic problem that shall be considered in the next section, the Newtonian gravitational potential of one or multiple \textit{rods} (rather than point-like masses) of mass $M$, infinitesimal width and mass density $\varrho$ plays an important role, since it is a key ingredient in the construction of Weyl solutions describing one or multiple black holes. Let us remark that, as a Newtonian problem, the passage from point particles to two equal rods, aligned with the direction of separation, does not alter the previous conclusion: in the symmetry plane, circular orbits still exist at any radial distance from the symmetry axis.

\section{The relativistic version}
\subsection{Uncharged case}
Weyl geometries \cite{weyl,exact} in four space-time dimensions admit two commuting, mutually orthogonal Killing vector fields. In vacuum, the metric may always be written in the form
\bequ
ds^2=-e^{2U(\rho,z)}dt^2+e^{-2U(\rho,z)}\left[e^{2k(\rho,z)}\left( d\rho^2+dz^2\right)+\rho^2d\phi^2\right] \ . \label{metric} \eequ
The vacuum Einstein equations, $R_{\mu\nu}=0$, reduce to a harmonic equation in an auxiliary Euclidean three space, $\mathbb{E}^3$, in cylindrical coordinates $(\rho,z,\phi)$,
\bequ
\Delta_{\mathbb{E}^3}U=0 \ , \label{Udifeq} \eequ
and two partial differential equations for $k(\rho,z)$
\bequ
\partial_z k=2\rho \partial_\rho U \partial_z U \ , \qquad \partial_\rho k=\rho[(\partial_\rho U)^2- (\partial_z U)^2] \ , \label{kdifeq} \eequ
which become line integrals once the solution for $U(\rho,z)$ is known.

A solution with two black holes is obtained by taking  $U(\rho,z)$ to be the potential of two finite rods of zero width and linear density $\varrho=1/2$.\footnote{This particular choice is required to avoid curvature singularities at the rod positions.} The rods are placed at $\rho=0$ and in the intervals $z\in [a_1,a_2]$ and  $z\in [a_3,a_4]$. Then
\bequ
e^{2U(\rho,z)}=\frac{(R_1-\zeta_1)(R_3-\zeta_3)}{(R_2-\zeta_2)(R_4-\zeta_4)} \ , \label{U2}\eequ
where we have introduced the quantities
\bequ
R_k\equiv \sqrt{\rho^2+\zeta_k^2} \ , \qquad \zeta_k\equiv z-a_k \ . \eequ
Introducing further 
\bequ
Y_{ij}\equiv R_iR_j+\zeta_i\zeta_j+\rho^2 \ , \eequ
the solution of \eqref{kdifeq}, given \eqref{U2}, is 
\bequ
e^{2k(\rho,z)}=\frac{Y_{43}Y_{21}Y_{41}Y_{32}}{4Y_{42}Y_{31}R_1R_2R_3R_4} \ . \label{k2} \eequ
An integration constant could be added to $k$. We have chosen this constant to be zero. With this choice, identifying the azimuthal coordinate with standard period $\phi\sim \phi +2\pi$, the metric is smooth, on the symmetry axis for $z<a_1$ and $z>a_4$. For $\rho=0$ and $a_2<z<a_3$ there is a conical excess, given by
\bequ
\delta=2\pi\left(e^{-k(\rho=0,a_2<z<a_3)}-1\right)=2\pi\frac{(a_4-a_3)(a_2-a_1)}{(a_4-a_1)(a_3-a_2)} \ . 
\eequ

The vacuum solution described by \eqref{metric}, \eqref{U2} and \eqref{k2} is actually a three parameters family of solutions. Physically, the three parameters may be taken to be the two black holes masses and the distance between them. The black hole masses may be computed as Komar integrals:
\bequ
M_1=\frac{1}{8\pi}\int_{S_1}\star d \xi=\frac{a_2-a_1}{2} \ , \qquad M_2=\frac{1}{8\pi}\int_{S_2}\star d \xi=\frac{a_4-a_3}{2} \ , 
\eequ 
where $\xi$ is the dual 1-form to the time-like Killing vector field $\partial/\partial t$ and the 2-surfaces $S_i$ correspond to the location of the two finite rods in Weyl coordinates. An asymptotic expansion shows that the ADM mass is
\bequ M_{ADM}=M_1+M_2 \ . \eequ
For the distance we shall take the coordinate distance in Weyl coordinates
\bequ
2L=a_3-a_2 \ . \eequ
This is a monotonic function of the proper distance
\bequ
\Delta z=\int_{a_2}^{a_3}\sqrt{g_{zz}}dz \ , \eequ
and hence a good measure thereof.

In order to have a totally geodesic sub-manifold we require $M_1=M_2\equiv M_K=M_{ADM}/2$. We also choose a symmetric coordinate system: $a_2=-L$ and $a_3=L$. 
The proper distance between the two black holes may then be expressed as
\bequ
\Delta z=\frac{8L(M_{ADM}+L)^2}{(M_{ADM}+2L)^2}E\left(\frac{L}{L+M_{ADM}}\right) , \eequ
where $E(x)$ is a complete elliptic integral of the second kind. It follows that $z=0$ is a totally geodesic sub-manifold, which has induced metric given by \eqref{metric} with $z=0$. Geodesics in this sub-manifold obey 
\begin{equation}
e^{2k(\rho,0)}\dot{\rho}^2=E^2-\left(e^{2U(\rho,0)}m^2+e^{4U(\rho,0)}\frac{J^2}{\rho^2}\right) \ , \label{geoequ} 
\end{equation}
where the dot represents derivative with respect to an affine parameter and we have introduced the energy $E$ and angular momentum $J$ for a particle of mass $m$.

\subsubsection{Circular photon orbits}
\begin{figure}[h!]
\centering\includegraphics[height=3.1in]{{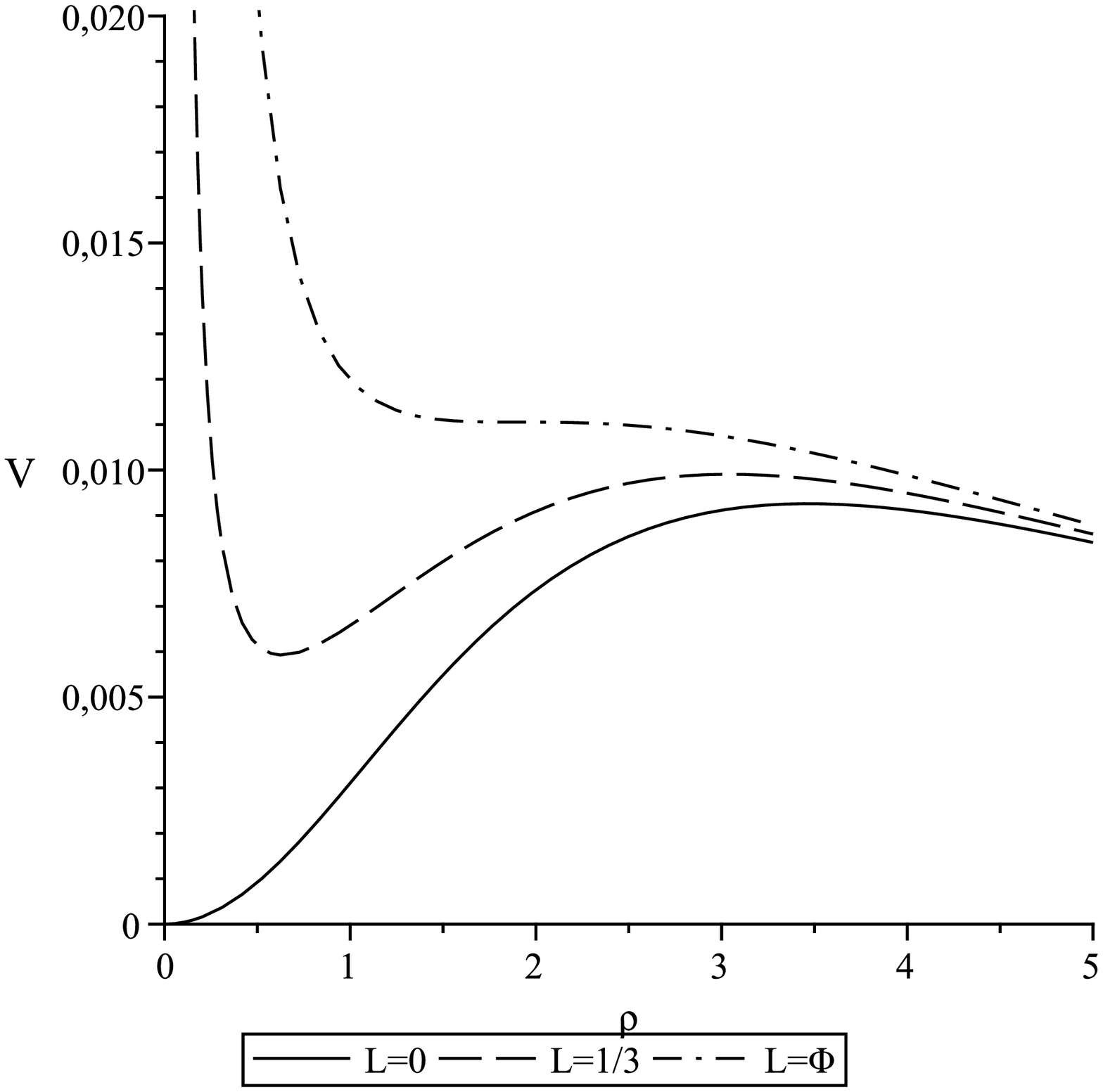}}
\centering\includegraphics[height=3.1in]{{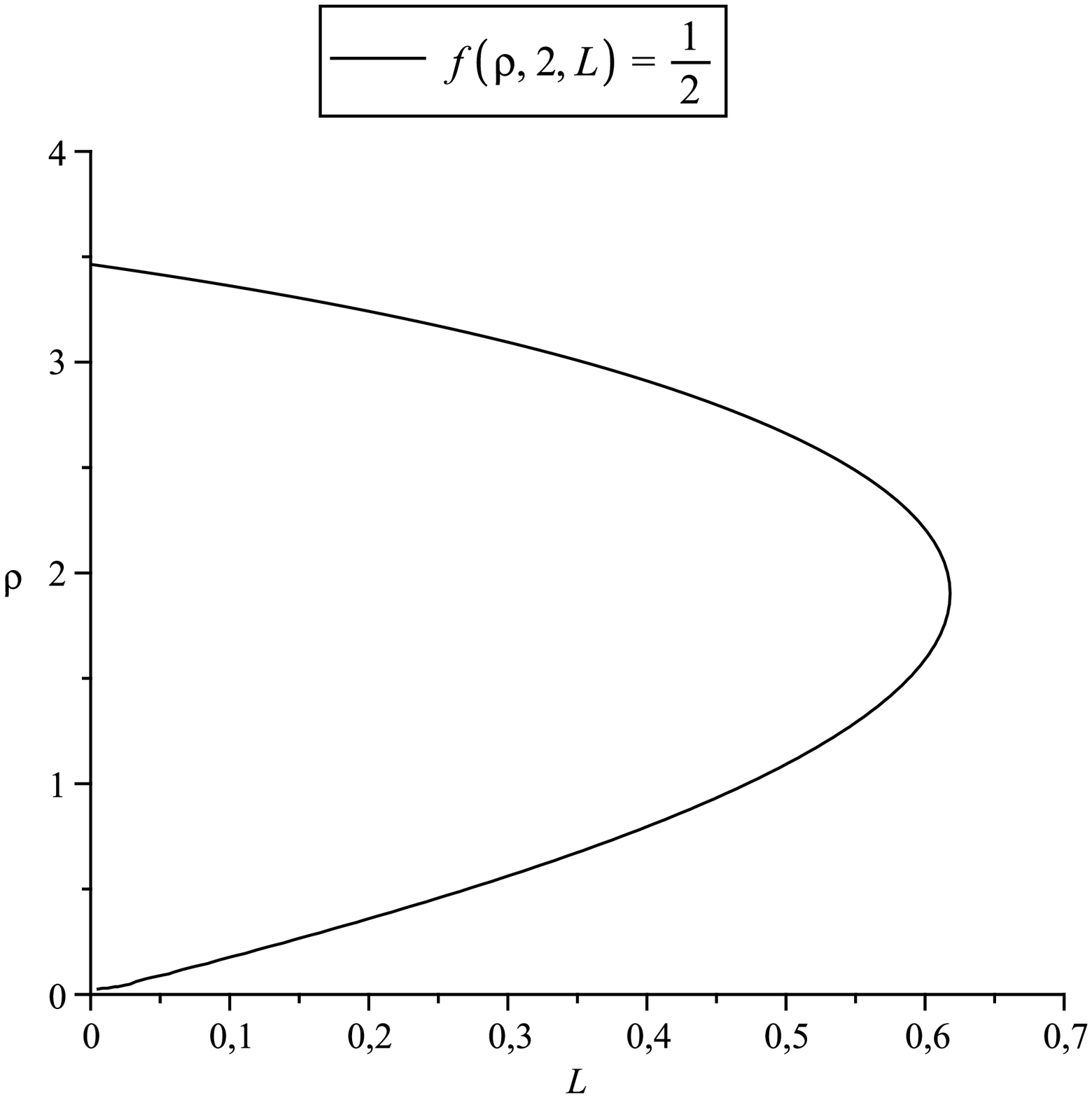}}
\begin{picture}(0,0)(0,0)
\end{picture}
\caption{Left: Effective potential for i) $L=0$, which has two extrema at $\rho_{CNO_1}(L=0)=0$ and $\rho_{CNO_2}(L=0)=\sqrt{3}M_{ADM}$; ii) $L=M_K/3$, for which the two extrema have moved closer to one another; iii) $L=L_{max}=\Phi M_K=M_K/\varphi $, for which the two extrema have merged. We have fixed $M_K=1$ in the plots. Right: Radial coordinate $\rho$ of the photon circular orbits as a function of $L$.}
\label{potential}
\end{figure}

From \eqref{geoequ}, circular null geodesics are determined by the extrema of the potential
\bequ
V(\rho)=\frac{e^{4U(\rho,0)}}{\rho^2}=\frac{1}{\rho^2}\left(\frac{L+\sqrt{L^2+\rho^2}}{M_{ADM}+L+\sqrt{(M_{ADM}+L)^2+\rho^2}}\right)^4 \ . \label{potentialV} \eequ
For $L=0$, i.e the single black hole limit, making the coordinate transformation
\bequ
\rho=r\sqrt{1-\frac{2M_{ADM}}{r}} \ , \eequ
the potential becomes
\bequ
V(r)=\frac{1}{r^2}\left(1-\frac{2M_{ADM}}{r}\right) \ . \label{potschwarzschild} \eequ
This is the effective potential for photons in a Schwarzschild black hole and in Schwarzschild coordinates. It has the well known extremum at $r=3M_{ADM}$. Another solution for a constant $r$ photon orbit is $r=2M_{ADM}$, corresponding to the null geodesic generator of the horizon. Thus, in Weyl coordinates we have two extrema, for $L=0$, at 
\bequ \rho_{CNO_1}(L=0)=0 \ , \qquad \rho_{CNO_2}(L=0)=\sqrt{3}M_{ADM} \ . \label{extrema}\eequ
For $L\gg M_{ADM}$, $V(\rho)$ has no extrema, since it is approximately  $V(\rho)\sim 1/\rho^2$. The interpolation between these two behaviours is as follows -  Fig \ref{potential}. As $L$ grows, the two extrema \eqref{extrema} approach one another, in both coordinate and proper distance, and merge for $L=L_{max}$.  Beyond this value of $L$, $V(\rho)$ has no extrema. To see the value of $L_{max}$, observe that the extrema of \eqref{potentialV} are given by
\bequ
f(\rho,M_{ADM},L)=\frac{1}{2} \ , \qquad  f(\rho,M_{ADM},L)\equiv \frac{M_{ADM}+L}{\sqrt{(M_{ADM}+L)^2+\rho^2}}-\frac{L}{\sqrt{L^2+\rho^2}} \ . \eequ
To solve for $\rho$ it is convenient to introduce  $x^2\equiv \rho^2/(M_{ADM}+L)^2$ and $y \equiv M_{ADM}/L$. Then the last equation becomes
\bequ
\frac{1}{2}=\frac{1}{\sqrt{1+x^2}}-\frac{1}{\sqrt{1+x^2(1+y)^2}} \ . \label{curve} \eequ
Take \eqref{curve} to define a curve $y=y(x)$, between $x=0$ and $x=\sqrt{3}$; at these values of $x$, $y\rightarrow \infty$. These two ``points" of the curve correspond to $\rho_{CNO_1}$ and $\rho_{CNO_2}$ given in \eqref{extrema}. Constant $y$ slices of the curve have two solutions for $y>y_{min}$, one solution for $y=y_{min}$ and no solution for $y<y_{min}$. To determine $y_{min}$, it is convenient to introduce
\bequ
\beta\equiv (1+y_{min})^{2/3} \ . \eequ 
Then, differentiating \eqref{curve}, extremising $y$ and replacing back into \eqref{curve} gives the quadratic equation
\bequ
\beta^2-3\beta+1=0 \ , \eequ
whose only solution leading to a positive $y_{min}$ is 
\bequ
\beta=1+\varphi \ , \qquad \varphi\equiv \frac{\sqrt{5}+1}{2} \ . \eequ 
$\varphi$ is the golden ratio. Using the recurrence property of the golden ratio 
\bequ
\varphi^{n+1}=\varphi^n+\varphi^{n-1} \ , \eequ
we arrive at $y_{min}=2\varphi$, and therefore at
\bequ
M_K=\varphi L_{max} \ \Leftrightarrow \  L_{max}=\Phi M_K \ , \qquad \Phi\equiv \frac{\sqrt{5}-1}{2} \ , \eequ
where $\Phi$ is the golden ratio conjugate. Thus, quite strikingly, the two null circular orbits merge when the ratio between the Komar mass of either black hole and the semi-distance between them is the golden ratio! This merging happens for Weyl radial coordinate 
\bequ
\rho=\sqrt{2+\varphi}M_K \ . \eequ

\subsubsection{Circular time-like orbits}
\begin{figure}[h!]
\centering\includegraphics[height=3.5in]{{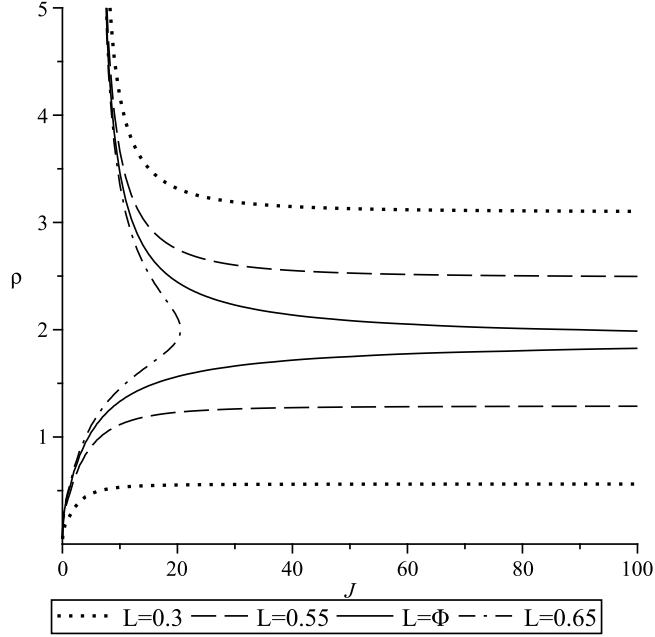}}
\begin{picture}(0,0)(0,0)
\end{picture}
\caption{Radial coordinate $\rho$ of the circular time-like orbit as a function of the angular momentum per unit mass $J$, for various values of $L$ and fixing $M_K=1$. One observes that there is a forbidden region $\rho\in [\rho_{CNO_1},\rho_{CNO_2}]$, for $L\le L_{max}$, where $\rho_{CNO_i}$ are the radii of the two circular null orbits. The time-like orbits approach the photon orbits for $J\rightarrow \infty$, from either side.}
\label{rhoj}
\end{figure}

\begin{figure}[h!]
\centering\includegraphics[height=3.8in]{{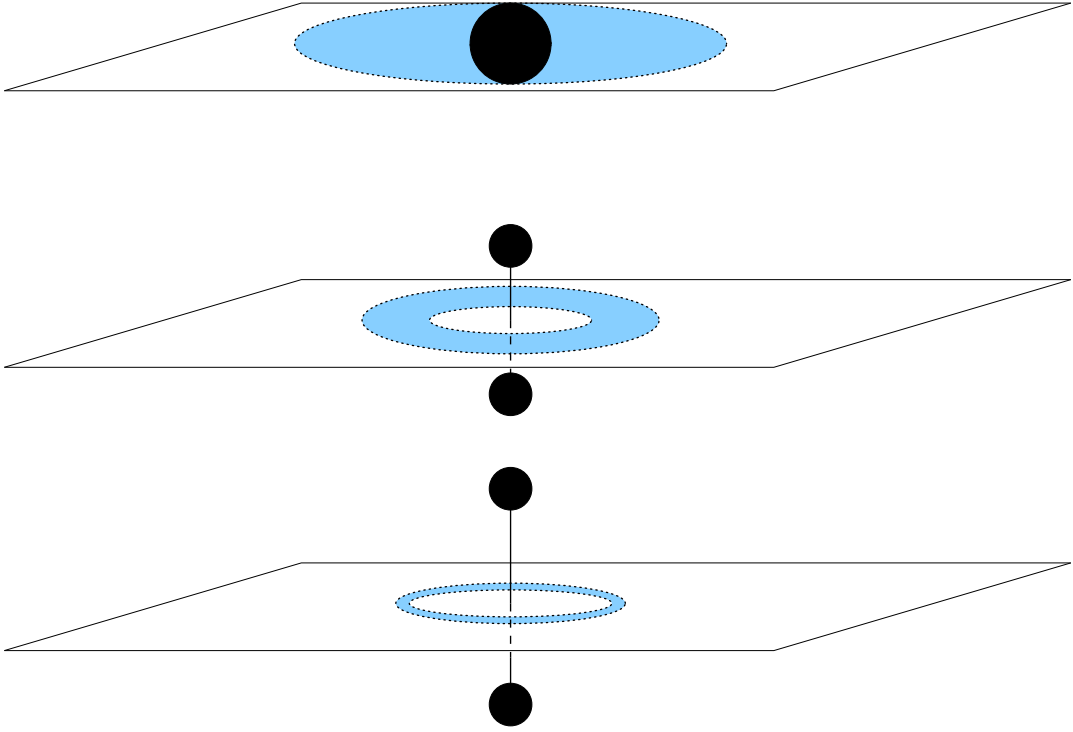}}
\begin{picture}(0,0)(0,0)
\end{picture}
\caption{Illustration of the behaviour of circular photon orbits  (dotted lines) in the double-Schwarzschild solution. For $L=0$, the two circular photon orbits sit at $r=3M_{ADM}$ (in Schwarzschild coordinates) and at the horizon (top). As $L$ increases these two photon orbits approach one another (middle) and merge as $L=\Phi M_K$ is approached (bottom). The two photon orbits delimit the forbidden annular region for circular time-like orbits (coloured region).}
\label{orbits}
\end{figure}

The significance of the circular photon orbits described in the last subsection can be seen by considering circular time-like orbits. From \eqref{geoequ} these are obtained by extremising the potential
\bea
V(\rho) & = &  e^{2U(\rho,0)}+J^2\frac{e^{4U(\rho,0)}}{\rho^2} \\ & = & \left(\frac{L+\sqrt{L^2+\rho^2}}{M_{ADM}+L+\sqrt{(M_{ADM}+L)^2+\rho^2}}\right)^2+ \frac{J^2}{\rho^2}\left(\frac{L+\sqrt{L^2+\rho^2}}{M_{ADM}+L+\sqrt{(M_{ADM}+L)^2+\rho^2}}\right)^4\ , \nonumber \eea
where $J$ is now the angular momentum per unit mass. The extrema of this potential are given by
\bequ
f(\rho,M_{ADM},L)=\frac{1}{2+(J\rho)^{-2}\left[\sqrt{(M_{ADM}+L)^2+\rho^2}+M_{ADM}+L\right]^2\left[\sqrt{L^2+\rho^2}-L\right]^2} \ . \label{timeuncharged}\eequ
In Fig. \ref{rhoj} we display the solution of Eq. \eqref{timeuncharged} for $\rho$, in terms of $J$, for fixed values of $L$. One observes that for $L<L_{max}=1/\varphi$ there is a \textit{forbidden band} for time-like circular geodesics, which is the region between the two photon circular orbits. This behaviour should be contrasted with that exhibited in Fig. \ref{newtonian} for the Newtonian case and is illustrated in Fig. \ref{orbits}.

\subsection{Charged case}
Let us start by recalling that in a Reissner-Nordstr\"om black hole, the circular null orbits are found as extrema of the potential
\bequ
V(r)=\frac{1}{r^2}\left(1-\frac{2M_{ADM}}{r}+\frac{q^2M_{ADM}^2}{r^2}\right) \ , \eequ
in standard Schwarzschild-type coordinates, where $q$ is the charge to mass ratio of the black hole. Extremising this potential gives the location of the circular null orbit outside the horizon
\bequ
r_{CNO_2}=\frac{3M_{ADM}}{2}+\sqrt{\left(\frac{3M_{ADM}}{2}\right)^2-2q^2M_{ADM}^2} \ , 
\eequ
whereas the second null orbit, with constant radial coordinate, relevant for our analysis is located at the horizon
\bequ
r_{CNO_1}=M_{ADM}+\sqrt{M_{ADM}^2-q^2M_{ADM}^2} \ . 
\eequ
Thus, the forbidden band for circular time-like orbits exists for all black holes in the Reissner-Nordstr\"om family. In the extremal case, this band acquires the special property of having infinite area, since the horizon, at $r=M_{ADM}=r_{CNO_1}$, is at an infinite proper radial distance from any point with radial coordinate $r>M_{ADM}$, which is the case for $r=2M_{ADM}=r_{CNO_2}$.

In order to consider two charged Reissner-Nordstr\"om black holes we shall now discuss charged Weyl solutions. The background fields are
\bequ
ds^2=-e^{2\bar{U}(\rho,z)}dt^2+e^{-2\bar{U}(\rho,z)}\left[e^{2k(\rho,z)}\left( d\rho^2+dz^2\right)+\rho^2d\phi^2\right] \ , \qquad A=-\chi(\rho,z)dt \ . \label{metriccharged} \eequ
The electrovacuum Einstein-Maxwell equations
\bequ
R_{\mu\nu}=2\left(F_{\mu\alpha}F_{\nu}^{\ \alpha}-\frac{1}{4}g_{\mu\nu}F_{\alpha\beta}F^{\alpha\beta}\right) \ , \qquad D_{\mu}F^{\mu\nu}=0 \ , \eequ
give rise to the set of equations 
\bequ
\Delta_{\mathbb{E}^3}\bar{U}=e^{-2\bar{U}}\left[(\partial_\rho \chi)^2+(\partial_z \chi)^2\right] \ , \qquad \Delta_{\mathbb{E}^3}\chi=2\left[\partial_\rho \chi \partial_\rho \bar{U}+\partial_z \chi\partial_z \bar{U}\right]  \ , \eequ
and
\bequ
\partial_z k=2\rho \partial_\rho \bar{U} \partial_z \bar{U}-2\rho e^{-2\bar{U}}\partial_\rho \chi \partial_z\chi \ , \qquad \partial_\rho k=\rho[(\partial_\rho \bar{U})^2- (\partial_z \bar{U})^2] -\rho e^{-2\bar{U}}[(\partial_\rho \chi)^2- (\partial_z \chi)^2] \ . \label{kdifeq2} \eequ
This set of equations is more involved than eq. \eqref{Udifeq}-\eqref{kdifeq}, in particular because there is no longer a linear equation to be solved. However, if one takes
\bequ
e^{2\bar{U}(\rho,z)}=1-\frac{2}{q}\chi(\rho,z)+\chi(\rho,z)^2 \ , \eequ
where $q$ is a constant, and 
\bequ
\chi(\rho,z)=\frac{q(1-e^{2U(\rho,z)})}{1-e^{2U(\rho,z)}+\sqrt{1-q^2}(1+e^{2U(\rho,z)})} \ , \eequ
the Einstein-Maxwell equations for \eqref{metriccharged} reduce to exactly the eq. \eqref{Udifeq}-\eqref{kdifeq} \cite{Azuma}. Thus we take $U(\rho,z)$ and $k(\rho,z)$ to be the same as in the uncharged case, \eqref{U2} and \eqref{k2}. The background \eqref{metriccharged} will then describe a double Reissner-Nordstr\"om solution, wherein both black holes have the same charge to mass ratio $q$.

The study of circular null geodesics will follow that of the uncharged case, \textit{mutatis mutandis}; they will be given by the extrema of the potential
\bequ
V(\rho)=\frac{e^{4\bar{U}(\rho,0)}}{\rho^2}=\frac{e^{4{U}(\rho,0)}}{\rho^2}\left[\frac{2\sqrt{1-q^2}}{1-e^{2{U}(\rho,0)}+\sqrt{1-q^2}(1+e^{2{U}(\rho,0)})}\right]^4 \ , \eequ
where $e^{{U}(\rho,0)}$ may be read off from \eqref{potentialV}. Such extrema obey
\bequ
f(\rho, M_{ADM},L)=\frac{1}{2}-\frac{Q}{Q+e^{-2U(\rho,0)}} \ , \eequ
where 
\bequ
Q\equiv \frac{1-\sqrt{1-q^2}}{1+\sqrt{1-q^2}} \ . \label{newcharge}\eequ

\begin{figure}[h!]
\centering\includegraphics[height=3.5in]{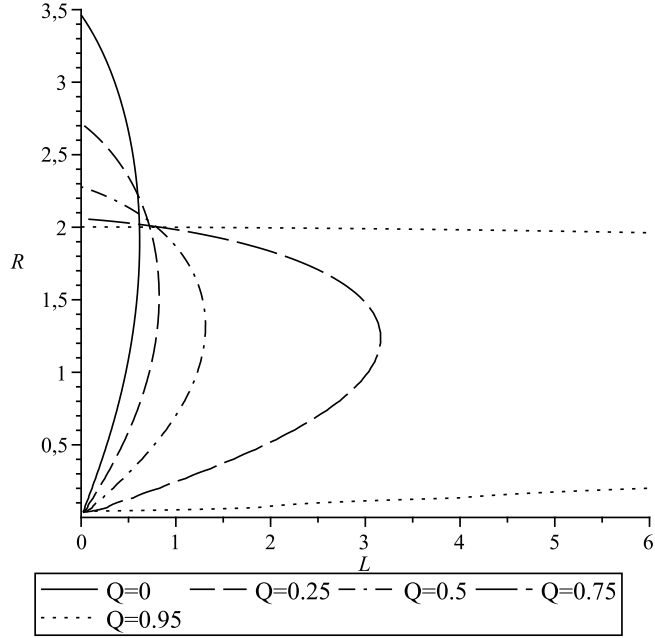}
\begin{picture}(0,0)(0,0)
\end{picture}
\caption{Radial coordinate of the photon circular orbits, $R=\rho \sqrt{1-q^2}$, in terms of the distance $L$ for various values of the charge  $Q$ defined by \eqref{newcharge}.}
\label{chargedplot}
\end{figure}

\begin{figure}[h!]
\centering\includegraphics[height=3.5in]{{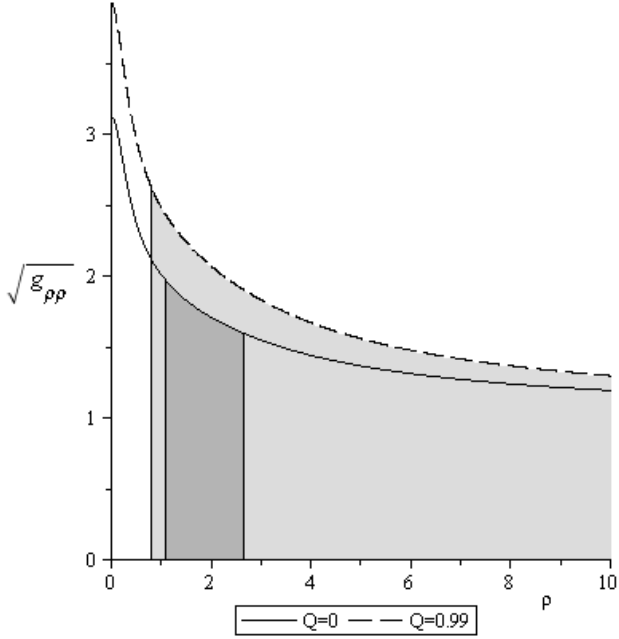}}
\begin{picture}(0,0)(0,0)
\end{picture}
\caption{$\sqrt{g_{\rho\rho}}$ as a function of $\rho$, for fixed $M_K=1$, $L=0.5$, and two values of the charge. The vertical lines correspond to the radii of the null circular orbits, and the shaded region corresponds to the area of the forbidden band for time-like circular orbits divided by $2\pi$. In the $Q=0.99$ case the exterior circular photon orbit is outside the $\rho$ range. Observe that the area increases with the charge.}
\label{areas}
\end{figure}

The result for the radii of the photon circular orbits in terms of the distance $L$ for various values of the charge is shown in Fig. \ref{chargedplot}, where we used the coordinate $R\equiv \rho \sqrt{1-q^2}$. The first feature we wish to emphasise is that, as $L$ is increased for fixed charge, the two photon orbits approach one another, just as for the uncharged case, and coalesce for a maximum value of $L$. Thus, the forbidden band for time-like circular orbits exists for all possible values of the charge, for sufficiently small $L$. The second feature we want to mention is that, fixing $L,M_K$ and increasing the charge, the forbidden band has larger area. For $L=0$ this area diverges as extremality is approached, as mentioned above. In Fig. \ref{areas} we exhibit an example with $L\neq 0$.

  In the extremal case, for which the solution is completely regular on and outside the event horizon, the analysis may also be done using the Majumdar-Papapetrou \cite{Majumdar,Papapetrou} form of the solution:
\bequ
ds^2=-\frac{dt^2}{H(\rho,z)^2}+H(\rho,z)^2(d\rho^2+\rho^2d\phi^2+dz^2) \ , \qquad A=-\frac{dt}{H(\rho,z)} \ , \eequ
where
\bequ
H(\rho,z)=1+\frac{M_K}{\sqrt{\rho^2+(z-L)^2}}+\frac{M_K}{\sqrt{\rho^2+(z+L)^2}} \  . \eequ
In this extremal case, circular null orbits exist in the $z=0$ sub-manifold as long as the equation
\bequ
(\rho^2+L^2)^{3/2}=2M_K(\rho^2-L^2) \ ,
\eequ
has solutions for real and positive $\rho$, which for fixed $M_K$ is possible when $L$ obeys
\bequ
L\le L_{max}\equiv \left(\frac{2}{3}\right)^{3/2}M_K \ . \eequ

\section{Interpretation: Optical Geometry}
A $D$ dimensional static geometry with $SO(D-1)$ isometry group may always be expressed by the line element
\bequ
ds^2=g_{tt}(R)dt^2+g_{RR}(R)dR^2+g_{\theta\theta}(R)d\Omega_{D-2} \ , \label{general} \eequ
where $d\Omega_{D-2}$ is the line element on the $(D-2)$-sphere. The optical geometry is the effective spatial geometry seen by light rays:
\bequ
ds^2_{\rm optical}=\frac{g_{RR}(R)}{|g_{tt}(R)|}dR^2+\frac{g_{\theta\theta}(R)}{|g_{tt}(R)|}d\Omega_{D-2} \ ,
\eequ
defined where $g_{tt}(R)<0$.

For the spacetime \eqref{general}, the radial equation of motion of a particle with mass, energy and angular momentum $m,E,J$, respectively is
\bequ
|g_{tt}(R)|g_{RR}(R)\dot{R}^2=E^2-|g_{tt}(R)|\left(m^2+\frac{J^2}{g_{\theta\theta}(R)}\right) \ . \label{orbitsstatic}\eequ
Circular null orbits are therefore determined by the extrema of the potential
\bequ
V(R)=\frac{|g_{tt}(R)|}{g_{\theta\theta}(R)} \ , \eequ
which is both the coefficient of the angular momentum term in \eqref{orbitsstatic} and the inverse of the proper radius squared of the sphere line element in the optical geometry. This potential pushes the test particle in the direction of increasing proper size of the spheres in the optical geometry. We dub this direction as \textit{outwards}. If it coincides with our na\"ive notion of ``outwards'', the angular momentum term may still be interpreted as a  \textit{centrifugal} term, as in flat space. Time-like circular orbits are only possible in a region where this term and the mass term in \eqref{orbitsstatic} originate forces in opposite directions.

\begin{figure}[h!]
\centering\includegraphics[height=3in]{{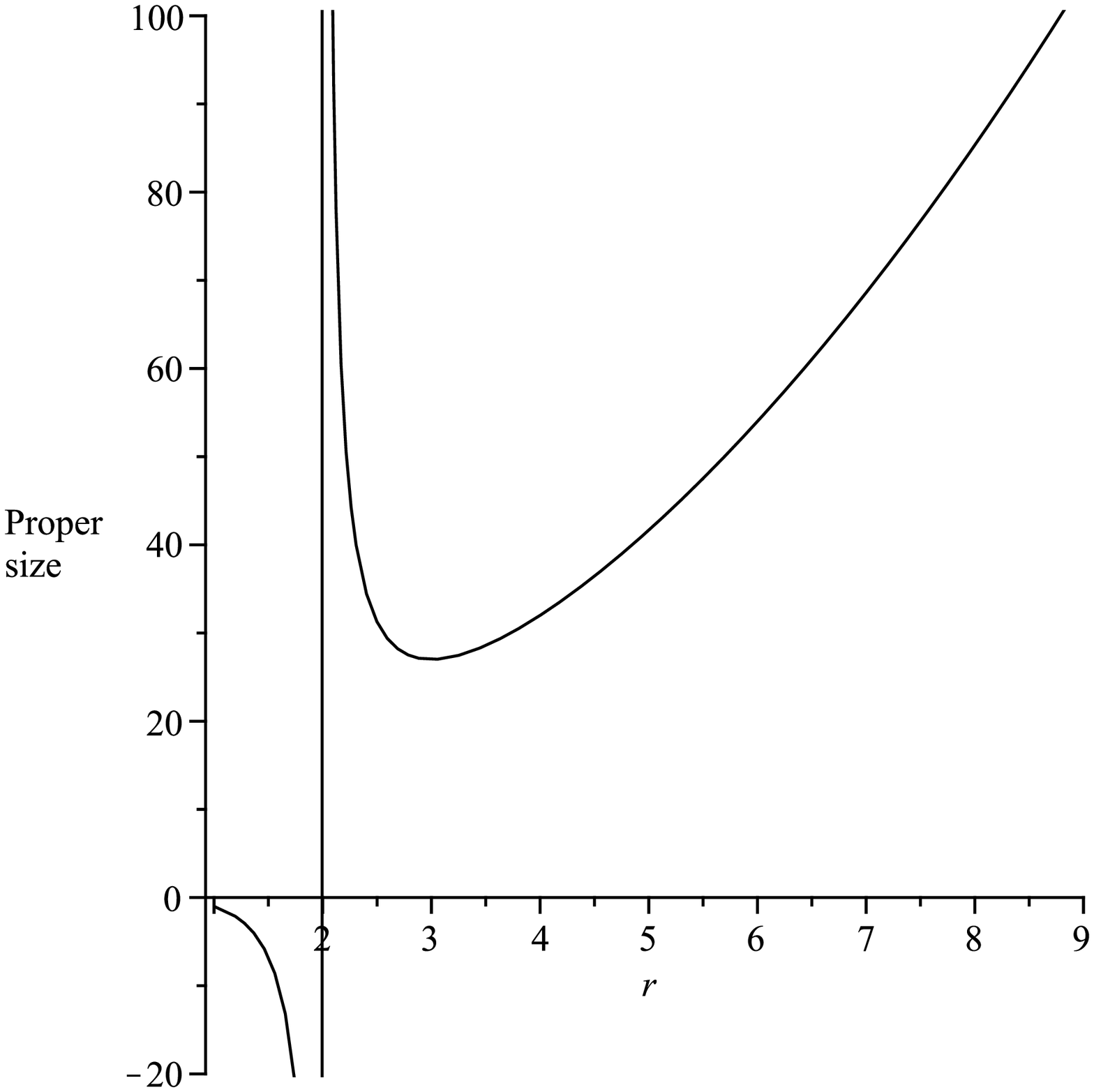}}
\centering\includegraphics[height=3in]{{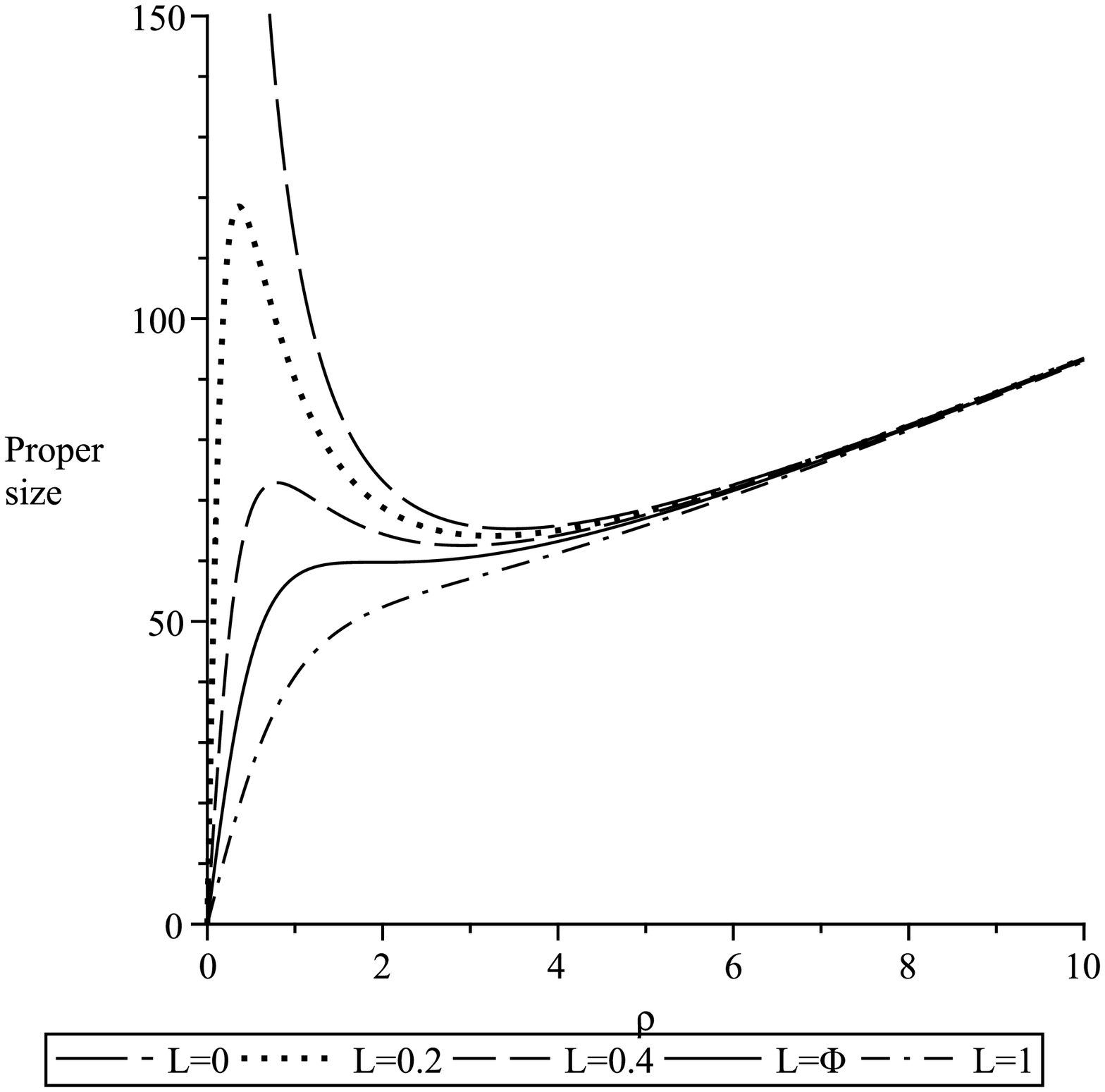}}
\begin{picture}(0,0)(0,0)
\end{picture}
\caption{Left (right): proper size of the 2-sphere (circle) in the optical geometry of the single (double) Schwarzschild solution. Defining ``outwards" as the direction in which this size increases, then ``outwards" corresponds to decreasing the radial coordinate for $r<3M_{ADM}$ (an annulus-like region, when $0<L<\Phi$). In the single Schwarzschild case, the optical metric is singular at $r=2M$; for $r<2M$, the proper size is negative since $V(r)<0$. In the double-Schwarzschild case, the annulus-like region lies in between the two extrema of the curves. We have set the individual black hole masses to unity.}
\label{optical1}
\end{figure}

For a single Schwarzschild black hole of mass $M$, circular time-like orbits are possible for $r>3M$ in Schwarzschild coordinates. Thus they are forbidden, outside the horizon, for $2M\le r\le 3M$. In this case, the optical geometry is
\bequ
ds^2_{\rm optical}=\frac{dr^2}{(1-2M/r)^2}+\frac{r^2}{1-2M/r}d\Omega_2 \ . \eequ
The proper size (area) of the 2-sphere is therefore $\mathcal{A}=4\pi / V(r)$, where the potential $V(r)$ is given by \eqref{potschwarzschild}. The area increases with $r$ for $r> 3M$, but decreases for $2M\le r\le 3M$ -  Fig. \ref{optical1} (left). Thus the ``centrifugal force" points towards the black hole, for  $2M\le r\le 3M$. Since the mass term is always attractive towards the black hole,  there can be no circular time-like orbits in the region $2M\le r\le 3M$.

For the double Schwarzschild solution the optical geometry in the symmetry plane is 
\bequ
ds^2_{\rm optical}=e^{-4U(\rho,0)}\left[e^{2k(\rho,0)} d\rho^2+\rho^2d\phi^2\right] \ . \label{metricoptical} \eequ
The proper size (perimeter) of the $\phi$ circle is therefore given by $\mathcal{L}=2\pi/ \sqrt{V(\rho)}$, where the potential $V(\rho)$ is given by \eqref{potentialV}.   The forbidden band for circular time-like orbits is the region wherein the radial coordinate decreases ``outwards", and therefore, the ``centrifugal force" becomes directed to the symmetry axis - Fig. \ref{optical1} (right) - preventing again the existence of time-like circular orbits.

\section{Multiple black holes}
\begin{figure}[h!]
\centering\includegraphics[height=3.0in]{{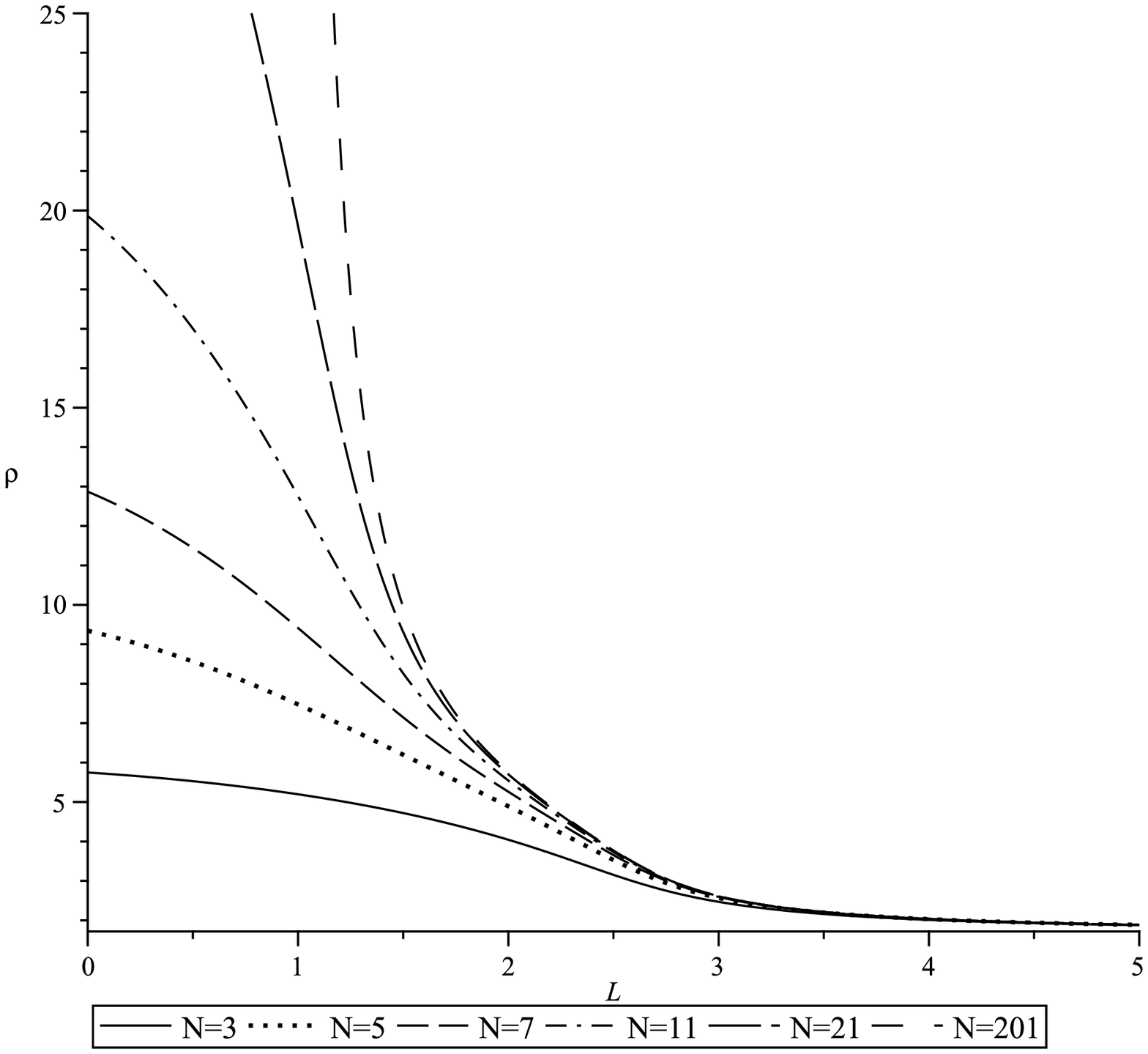}}
\centering\includegraphics[height=3.0in]{{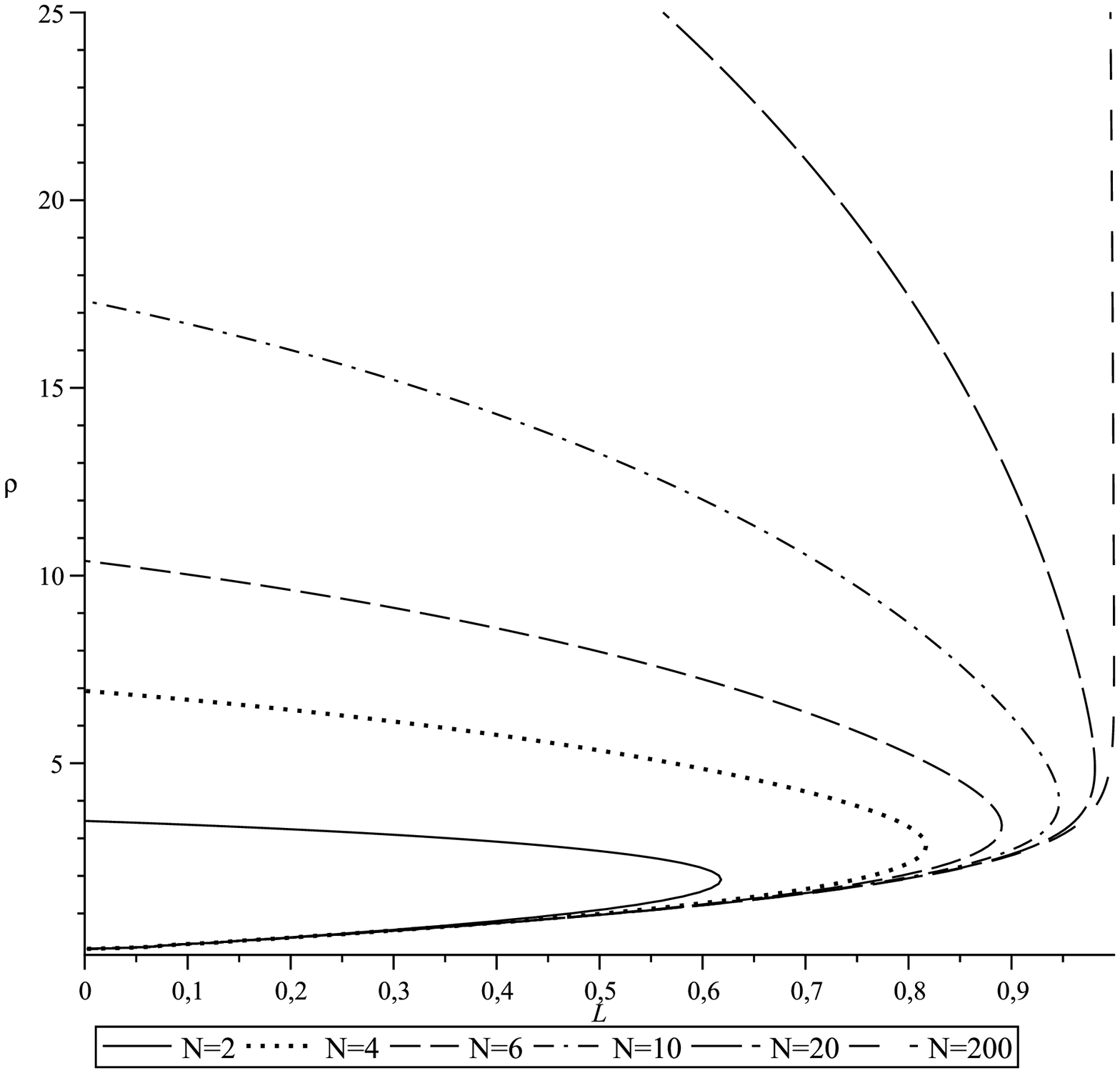}}
\begin{picture}(0,0)(0,0)
\end{picture}
\caption{Radial coordinate of circular null geodesics for an array of $N$ black holes as a function of $L$. Left (right): N odd (even).}
\label{multiple}
\end{figure}

We shall now consider a Weyl solution with $N$ uncharged black holes. Then \eqref{U2} is replaced by 
\bequ
e^{2U(\rho,z)}=\frac{(R_1-\zeta_1)(R_3-\zeta_3)}{(R_2-\zeta_2)(R_4-\zeta_4)} \dots \frac{(R_{2N-1}-\zeta_{2N-1})}{(R_{2N}-\zeta_{2N})} \ . \label{U2many}\eequ
This function $U(\rho,z)$ describes the Newtonian potential of $N$ rods of infinitesimal width and mass density $\varrho=1/2$, located at $\rho=0$ and with $z$ coordinate in the intervals, respectively,
\bequ
[a_1,a_2] \ , \qquad [a_3,a_4] \ , \qquad  \dots \ , \qquad [a_{2N-1},a_{2N}] \ . \eequ 
Imposing 
\bequ
a_n=-a_{2N+1-n} \ , \qquad n=1\dots 2N \ , \eequ
the geometry still admits a discrete $\mathbb{Z}_2$ symmetry, of which $z=0$ is a fixed point set. Then, the circular null orbits in this totally geodesic sub-manifold are obtained by extremising the potential
\bequ
V(\rho)=\frac{e^{4U(\rho,0)}}{\rho^2}=\frac{1}{\rho^2}\left(\frac{(\sqrt{\rho^2+a_1^2}+a_1)(\sqrt{\rho^2+a_3^2}+a_3)}{(\sqrt{\rho^2+a_2^2}+a_2)(\sqrt{\rho^2+a_4^2}+a_4)} \dots \frac{\left(\sqrt{\rho^2+a_{2N-1}^2}+a_{2N-1}\right)}{\left(\sqrt{\rho^2+a_{2N}^2}+a_{2N}\right)}\right)^2 \ ,
\eequ
which yields the condition
\bequ
\sum_{n=1}^{2N}\frac{(-1)^n a_n}{\sqrt{\rho^2+a_n^2}}=1 \ . \eequ
We set
\bequ
a_{2n}-a_{2n-1}=2M_K \ , \qquad a_{2n+1}-a_{2n}=2L \ , \qquad n=1\dots N \ , \eequ
such that the coordinate distance between each two black holes is $2L$ and their Komar mass is $M_K$. In Fig. \ref{multiple} we have plotted the radial coordinate of circular null orbits versus $L$, for various numbers of black holes. For $N$ even, the behaviour is quite similar to that discussed for $N=2$. Moreover 
\bequ
\lim_{N\rightarrow +\infty} L_{max}= M_K \ . \eequ
For $N$ odd, the behaviour is quite similar to that of a single Schwarzschild black hole, due to the existence of a black hole in the symmetry plane.

\section{Final Remarks}
In this paper we have studied a relativistic version of Euler's 3-body problem: the motion of test particles in a $\mathbb{Z}_2$ invariant Weyl solution describing two (or more) black holes. We have considered both uncharged and charged black holes, within Einstein-Maxwell theory. Generically, these solutions have conical singularities on the symmetry axis. In the extremal case, however, they become a special case of the Majumdar-Papapetrou spacetimes, and are completely regular on and outside the event horizon \cite{Hartle:1972ya,Gibbons:1980tf}.

For sufficiently small distance between the two black holes, there are two circular null geodesics in the symmetry plane corresponding to the set of fixed points of the $\mathbb{Z}_2$ symmetry. The interior one is stable (against perturbations in the symmetry plane), while the exterior one is unstable. These orbits delimit a forbidden region for circular time-like geodesics. In the optical geometry, the proper size of the $SO(2)$ isometry group orbits grows towards the symmetry axis, rather than away from it, in this region. 
This unusual property is correlated with a physical effect that may be, heuristically, described as  the centrifugal force becoming directed towards the symmetry axis. Thus, it cannot balance the gravitational attraction, preventing the existence of time-like circular orbits.

The existence of a stable circular null geodesic is associated with a behaviour which may be described as an `optical fibre' or `wave guide' like geometry. Photons with sufficiently low energy will be caught in a potential well in the radial direction, as may be observed in Fig. \ref{potential}. Thus, as they move in the $\phi$ direction, photons will oscillate between a maximum and a minimum radii. The wave guide like geometry may allow light waves to travel around the $\phi$ direction without decreasing their amplitude significantly.\footnote{In a very different context, this resembles the SOFAR (Sound Fixing and Ranging) channel \cite{gary}, in the ocean, which occurs at the depth where the sound velocity is minimal, and allows the singing of whales to travel very long distances \cite{synge}.}

As the distance between the two black holes is increased, the two null circular orbits approach one another, eventually coalescing. For the uncharged case this happens when the ratio between the Komar mass of each black hole and the semi-distance between the black holes is the golden ratio. 

One may ask whether the geodesic equations in the full spacetime, rather than just the symmetry plane, are Liouville integrable. In \cite{Cornish:1996de} a similar question was studied for extremal black holes in Einstein-Maxwell-dilaton theory, for arbitrary dilaton coupling. These authors showed that the geodesic motion is generically chaotic. There is only a very special case of integrable motion: when the dilaton coupling is that of Kaluza-Klein theory. For the generic case studied herein, of charged Weyl solutions that may be continuously connected to the extremal black holes of Einstein-Maxwell theory (with zero dilaton coupling therefore), it seems quite unlikely that the motion may be integrable. Observe however, that the uncharged double Schwarzschild solution may also be continuously connected to the extremal black holes of Kaluza-Klein theory. Thus, it would be interesting to understand where, along this trajectory in the space of solutions, integrability appears.

\section*{Acknowledgments}
We are very grateful to G. W. Gibbons and E. Teo for correspondence and comments on drafts of this paper. F.C. is supported by a junior research grant from Centro de F\'\i sica do Porto. C.H. is supported by a Ci\^encia 2007 research contract. This work as been further supported by the FCT grant CERN/FP/83508/2008.


\begin{thebibliography}{99}

\bibitem{Coulson}
C.~A.~Coulson and A.~Joseph,
``A constant of the motion for the two-centre Kepler problem,"
Int. Jour. Quant. Chem. {\bf 1} (1967) 337.

\bibitem{Bell}
D.~Lynden-Bell,
``A simple derivation and interpretation of the third integral in stellar dynamics,"
Mon. Not. R. Astron. Soc. {\bf 338} (2003) 208.

\bibitem{Gibbons:2006mi}
  G.~W.~Gibbons and C.~M.~Warnick,
  ``Hidden symmetry of hyperbolic monopole motion,''
  J.\ Geom.\ Phys.\  {\bf 57} (2007) 2286
  [arXiv:hep-th/0609051].

\bibitem{Will:2008ys}
  C.~M.~Will,
  ``Carter-like constants of the motion in Newtonian gravity and
  electrodynamics,''
  Phys.\ Rev.\ Lett.\  {\bf 102} (2009) 061101
  [arXiv:0812.0110 [gr-qc]].


\bibitem{Abram}
M.~A.~Abramowicz,
``Black Holes and the Centrifugal Force Paradox,"
Scientific American, March 1993, 74-81.

\bibitem{Gibbons:2008zi}
  G.~W.~Gibbons, C.~A.~R.~Herdeiro, C.~M.~Warnick and M.~C.~Werner,
  ``Stationary Metrics and Optical Zermelo-Randers-Finsler Geometry,''
  Phys.\ Rev.\  D {\bf 79} (2009) 044022
  [arXiv:0811.2877 [gr-qc]].
  
  \bibitem{jose}
  J.~V.~Jos\'e and E.~J.~Saletan,
``Classical Dynamics (a contemporary approach),''
Cambridge University Press, 1998.

\bibitem{weyl}
H.~Weyl, 
Ann. \ Phys. (Leipzig) {\bf 54} (1917) 117.

\bibitem{exact}
H.~Stephani et al.,
``Exact Solutions of Einstein's Field Equations,"
Cambridge Monographs on Mathematical Physics, 2003.


\bibitem{Azuma}
T.~Azuma and T.~Koikawa,
``Equilibrium condition in the axisymmetric N-Reissner-Nordstrom solution,''
Prog.~Theor.~Phys. {\bf 92} (1994) 1095.

\bibitem{Majumdar}
S.~D.~Majumdar,
Phys. Rev. {\bf 72} (1947) 390.

\bibitem{Papapetrou}
A.~Papapetrou,
Proc. R. Irish Acad. {\bf A51} (1947) 191.

\bibitem{Hartle:1972ya}
  J.~B.~Hartle and S.~W.~Hawking,
  ``Solutions of the Einstein-Maxwell equations with many black holes,''
  Commun.\ Math.\ Phys.\  {\bf 26} (1972) 87.
  
\bibitem{Gibbons:1980tf}
  G.~W.~Gibbons,
  ``Non-Existence Of Equilibrium Configurations Of Charged Black Holes,''
  Proc.\ Roy.\ Soc.\ Lond.\  A {\bf 372} (1980) 535.
  
  \bibitem{gary}
  G.~W.~Gibbons,
  Private communication.
  
  \bibitem{synge}
  C.~S.~Morawetz,
  ``Geometrical Optics and the singing of whales,"
  The Amer. Math. Mon. {\bf 85} (1978) 548.

\bibitem{Cornish:1996de}
  N.~J.~Cornish and G.~W.~Gibbons,
  ``The tale of two centres,''
  Class.\ Quant.\ Grav.\  {\bf 14} (1997) 1865
  [arXiv:gr-qc/9612060].

\end{thebibliography}
\end{document}